\documentclass[a4paper,11pt]{article}
\usepackage[reqno,fleqn]{amsmath}
\usepackage{color,graphicx}
\usepackage[T1]{fontenc}
\author{Behnam Mohammadi$^a$\footnote{be.mohammadi@urmia.ac.ir}, Elnaz Amirkhanlou$^b$\footnote{eliamirkhanlou@yahoo.com}\\
Department of Physics, Urmia University, Urmia, Iran}
\title{Estimated of $CP$ violation in $B^0$ meson decays into $D^{*+}$ and $D^-$ mesons}
\begin{document}
\maketitle
\begin{abstract}

The decay $B^0\rightarrow D^{*+}D^-$ is favorable mode for
studying $CP$ violation in the interference between mixing and
decay for $B^0$ and $\bar{B}^0$ mesons. The latest analysis of the
$CP$ parameters has been performed by the LHCb collaboration
values of $S_{D^*D}=-0.861\pm0.077\pm0.019$,
$C_{D^*D}=-0.059\pm0.092\pm0.020$, $\triangle
S_{D^*D}=0.019\pm0.075\pm0.012$, $\triangle
C_{D^*D}=-0.031\pm0.092\pm0.016$, and
$\mathcal{A}_{D^*D}^{CP}=0.008\pm0.014\pm0.006\pm0.003$. We have
been estimated the parameters $S_{D^*D}$ and $C_{D^*D}$ of the
$B^0\rightarrow D^{*+}D^-$ decay as $-0.709\pm0.024$ and
$-0.051\pm0.004$. In the following, we have obtained the values of
$\triangle S_{D^*D}=0.054\pm0.003$ and $\triangle
C_{D^*D}=0.020\pm0.001$ and direct $CP$ violation of
$0.008\pm0.001$. Also, we have calculated the branching ratio of
$B^0\rightarrow D^{*+}D^-$ decay.
The values obtained in this work are comparable with the corresponding experimental values.\\
\end{abstract}

\section{Introduction}
The standard model (SM) is a relativistic quantum field theory
that involves the search for fundamental particles and the
fundamental interactions that occurring among them. To perform
such searches through high-precision measurements of the
parameters of the quark-flavour of the SM sector with $b-$ and
$c-$hadron decays is developed. In this way, possible
inconsistencies with the SM predictions are revealed. The
increasing amount of data makes it necessary to consider
higher-order the SM corrections \cite{LHCb3}.\\
One way to do this is to examine decays that involve $b\rightarrow
c\bar{c}d$ transitions, such as $B^0\rightarrow D^{*+}D^-$.
Neutral meson mixing is one important effect that allows access to
parameters in the flavour sector \cite{Belle2}. The mesons
composed of a different quarks and anti-quarks type decay weakly,
allowing $CP$ violation and mixing. Mixing describes the
transformation of a neutral meson into an antiparticle state and
vice versa, and is also called meson oscillation. The
time-dependent oscillation between the particle and antiparticle
states appears \cite{AJ1}.\\
$CP$ violation in general could lead to the excess of a
matter-antimatter in our universe, but the smallness of the
observed $CP$ violation is not sufficient to explain the
observations \cite{PH1}. Nevertheless, the fact that the $CP$
violation is a relatively small non-zero value is interesting and
allows for further studies on its properties. Also, new sources of
$CP$ violation beyond the SM that account for the difference
between measured values and SM predictions can be considered as a
research idea for the yet-undiscovered physics \cite{ZH1}.\\
In the case of $CP$ symmetry in the $B$ meson system, we can study
the processes in which the $B$ mesons decay into a $CP$-eigenstate
state. In a general way, we can compare the rate at which a $B$
meson decays into a $CP$-eigenstate with the rate at which a $B$
meson decays into a $CP$-conjugate final state ($\bar{f}$), to the
rate at which a $\bar{B}$ meson decays into the $CP$ final state
($f$) and to the rate at which a $\bar{B}$ meson decays into the
$\bar{f}$. These different final states provide additional
information about the system, and only by combining such
information from different measurements can we get a complete
picture of the subject as well as accurate results. The difference
between the $B^0$ and $\bar{B}$ meson decays appears only in the
time-dependent decay rate, and this time corresponds to the time
when the $B$ meson freely propagates before it decays to the
$CP$-eigenstate \cite{HFLAV2}. In Tabs. \ref{tab1} and \ref{tab2}
an overview of existing measurements and the world average is
provided for the $B^0\rightarrow D^{*\pm}D^{\mp}$ decays by the
different
collaborations.\\
\begin{table}[t]
\centering\caption{\label{tab1} Experimentally values $CP$
violation parameters for $B^0\rightarrow D^{*\pm}D^{\mp}$ decay.}
\begin{tabular}{|c|c|c|c|c|}
  \hline
 & Belle \cite{Belle1} & BABAR \cite{BABAR1} & LHCb \cite{LHCb1} &
HFLAV \cite{HFLAV1} \\ \hline
 $S_{D^*D}$ & $-0.78\pm0.15\pm0.05$ & $-0.68\pm0.15\pm0.04$ &  $-0.861\pm0.077\pm0.019$ & $-0.73\pm0.11$ \\ \hline
 $C_{D^*D}$ & $-0.01\pm0.11\pm0.04$ & $+0.04\pm0.12\pm0.03$ & $-0.059\pm0.092\pm0.020$ & $0.01\pm0.09$ \\ \hline
 $\triangle S_{D^*D}$ & $-0.13\pm0.15\pm0.04$ & $+0.05\pm0.15\pm0.02$ & $+0.019\pm0.075\pm0.012$ & $-0.041\pm0.11$ \\ \hline
 $\triangle C_{D^*D}$ & $+0.12\pm0.11\pm0.03$ & $+0.04\pm0.12\pm0.03$ & $-0.031\pm0.092\pm0.016$ & $0.08\pm0.08$ \\ \hline
 $\mathcal{A}_{D^*D}^{CP}$ & $+0.06\pm0.05\pm0.02$ & $+0.008\pm0.048\pm0.013$ & $0.008\pm0.014\pm0.006$ & $0.03\pm0.04$ \\ \hline
\end{tabular}
\end{table}
\begin{table}[t]
\centering\caption{\label{tab2} Measured results of the
time-dependent $CP$ violatin parameters for $B^0\rightarrow
D^{*\pm}D^{\mp}$ decay.}
\begin{tabular}{|c|c|c|c|c|}
  \hline
$B^0\rightarrow D^{*\pm}D^{\mp}$ & BABAR \cite{BABAR2} & Belle
\cite{Belle2} & PDG [2021] . Ave \cite{PDG1} \\ \hline
 $S_{D^{*+}D^-}$ & $-0.82\pm0.75\pm0.14$ & $-0.55\pm0.39\pm0.12$ &  $-0.80\pm0.09$ \\ \hline
 $C_{D^{*+}D^-}$ & $-0.47\pm0.40\pm0.12$ & $-0.37\pm0.22\pm0.06$ & $-0.03\pm0.09$ \\ \hline
 $S_{D^{*-}D^+}$ & $-0.24\pm0.69\pm0.12$ & $-0.96\pm0.43\pm0.12$ & $-0.83\pm0.09$ \\ \hline
 $C_{D^{*-}D^+}$ & $-0.22\pm0.37\pm0.10$ & $+0.23\pm0.25\pm0.06$ & $-0.02\pm0.08$ \\ \hline
\end{tabular}
\end{table}
Recently, the first measurement of $CP$ violation in the
$B^0\rightarrow D^{*\pm}D^{\mp}$ decay has been reported in the
LHCb experiment. They have measured the $CP$ parameters as
$S_{D^*D}=-0.861\pm0.077\pm0.019$, $\triangle
S_{D^*D}=0.019\pm0.075\pm0.012$,
$C_{D^*D}=-0.059\pm0.092\pm0.020$, $\triangle
C_{D^*D}=-0.031\pm0.092\pm0.016$ and $\mathcal{A}_{D^*D}^{CP}=0.008\pm0.014\pm0.006\pm0.003$ \cite{LHCb1}.\\
In this work, we have estimated the $CP$ parameters and branching
ratio for the $B^0\rightarrow D^{*\pm}D^{\mp}$ decay. Under the
factorization approach, the amplitudes of $B^0\rightarrow
D^{*\pm}D^{\mp}$ decay can be obtained as separate factorizable
contributions that include the current-current and penguin
contributions. In the case of $\langle B^0\rightarrow
D^-\rangle\times\langle 0\rightarrow D^{*+}\rangle$ ($\langle
\bar{B}^0\rightarrow D^+\rangle\times\langle 0\rightarrow
D^{*-}\rangle$) where the matrix elements $B^0$ to $D^-$
($\bar{B}^0$ to
$D^+$) transition multiplying $D^{*+}$ ($D^{*-}$) arising from the vacuum.\\
We have obtained the branching fraction using the decay amplitude
that is to be $\mathcal{B}(B^0\rightarrow
D^{*+}D^-)=(5.20\pm1.25)\times10^{-4}$ at $\mu=2m_b$ scale. This
value is well compatible with the value of
$\mathcal{B}(B^0\rightarrow D^{*+}D^-)=(6.03\pm0.50)\times10^{-4}$
reported by HFLAV \cite{HFLAV1}.\\
We have estimated the $CP$ violation as
$\mathcal{A}_{D^*D}^{CP}=0.008\pm0.001$ and we have obtained other
parameters of $CP$ violation, such as $S_{D^*D}=-0.709\pm0.024$,
$\triangle S_{D^*D}=0.054\pm0.003$, $C_{D^*D}=-0.051\pm0.004$ and
$\triangle C_{D^*D}=0.020\pm0.001$.

\section{Branching fraction and $CP$ violation in $B^0\rightarrow D^{*+}D^-$ decay}

\begin{figure}[t]
\begin{center} \includegraphics[scale=0.9]{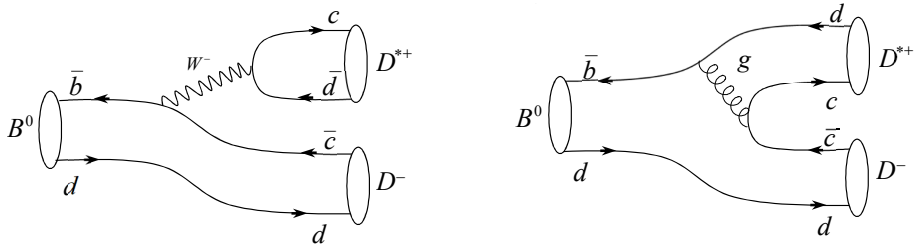}
\caption{\label{fig1}Feynman diagrams contributing to
$B^0\rightarrow D^{*+}D^-$ decay.}\end{center}
\end{figure}

We explained the theoretical background of $CP$ violation in the
$B^0$ meson system using the SM of particle physics and its
constructed theoretical framework. We then presented an overview
of the field of flavour physics, including the basic ideas of quark mixing and $CP$ violation in the $B$ meson.\\
Now we want to calculate the direct $CP$ violation. The direct
$CP$ violation arises in the ratio of the amplitude $A_f
(f=D^{*+}D^-)$ to its conjugate amplitude ($\bar{A}_{\bar{f}}
(\bar{f}=D^{*-}D^+))$. In this case, two types of phases occur in
these amplitudes \cite{A.I1}. The first type of phase is created
in complex parameters in the Lagrangian. In the SM, these phases
occur only in the CKM matrix and are called weak phases ($\phi_i$)
\cite{G.B1}. The CKM matrix elements are in the unitarity triangle
relation, $V_{ub}^*V_{uq}+V_{cb}^*V_{cq}+V_{tb}^*V_{tq}=0$
$(q=d,s)$ and
weak phases are introduced as $\phi_1={\rm{arg}}(V_{cq})$ and $\phi_2=-{\rm{arg}}(V^*_{tb})$.\\
Another type of phase can appear in the scattering or decay
amplitudes that are called the strong phases ($\delta_i$). these
phases occur even when the Lagrangian is real. Such phases do not
violate $CP$ because they appear in amplitudes ($A_f$ and
$\bar{A}_{\bar{f}}$) with the same sign. Their origin is the
possible contribution of the mode the intermediates on-shell
states in the process of decay. In fact, it is an absorptive part
of an amplitude that has contributions from coupled channels. The
dominant re-scattering is due to strong interactions and this is
the reason for naming these phases. The $CP$ violation will not
occur unless we have different strong phases in addition to
different weak phases \cite{G.B1}. The strong phase $\delta_1$ is
obtained from $|\mathcal {A}_1| e^{i\delta_1}$ and $\delta_2$ from
$|\mathcal {A}_2| e^{i\delta_2}$.\\
The $B^0\rightarrow D^{*+}D^-$ decay (and $\bar{B}^0\rightarrow
D^{*-}D^+$ decay), with two contributing amplitudes $\mathcal
{A}_1$ and $\mathcal {A}_2$. This means that the decay can be done
by two different paths and those are tree ($\mathcal {A}_1$) and
penguin ($\mathcal {A}_2$) diagrams. For the total decay amplitude
we have \cite{M.S.S}: {\setlength\arraycolsep{.75pt}
\begin{eqnarray}\label{eq1}
\mathcal {A}(B^0 \rightarrow D^{*+}D^-)&=&|\mathcal {A}_1|
e^{i\delta_1}e^{i\phi_1} +|\mathcal
{A}_2|e^{i\delta_2}e^{i\phi_2},
\end{eqnarray}}\label{eq2}
where $|\mathcal {A}_1|$ and $|\mathcal {A}_2|$ represent
$|\mathcal {A}_1(B^0\rightarrow D^{*+}D^-)|$ and $|\mathcal
{A}_2(B^0\rightarrow D^{*+}D^-)|$.\\
Feynman tree diagrams have the largest amplitude contribution
compared to penguin diagrams. The Feynman diagrams of
$B^0\rightarrow D^{*+}D^-$ decay are shown in Fig. \ref{fig1}
and the decay amplitude can be expressed as
{\setlength\arraycolsep{.75pt}
\begin{eqnarray}\label{eq3}
\mathcal{A}(B^0\rightarrow D^{*+}D^-)&=&\sqrt{2}iG_{F}f_{D^*}
F_1^{B \rightarrow D}(m_{D^*}^2)\Big(V_{cb}^*V_{cd}a_1 -
V_{tb}^*V_{td}(a_4 + a_{10} + (a_6 + a_8)r_\chi^{D^*})\Big),
\end{eqnarray}}
where the tree and penguin level amplitudes are as follows,
respectively{\setlength\arraycolsep{.75pt}
\begin{eqnarray}\label{eq4}
\mathcal{A}_1 (B^0\rightarrow D^{*+}D^-)&=&\sqrt{2}iG_{F}f_{D^*}
F_1^{B \rightarrow D}(m_{D^*}^2)V_{cb}^*V_{cd}a_1,
\end{eqnarray}}
and {\setlength\arraycolsep{.75pt}
\begin{eqnarray}\label{eq5}
\mathcal{A}_2 (B^0\rightarrow D^{*+}D^-)&=&\sqrt{2}iG_{F}f_{D^*}
F_1^{B \rightarrow D}(m_{D^*}^2) V_{tb}^*V_{td}\Big(a_4 + a_{10} +
(a_6 + a_8)r_\chi^{D^*}\Big),
\end{eqnarray}}
the quantity of $r_\chi^{D^*}$ is equal to
$(2m_{D^*}/m_b)(f_{D^*}^\perp/f_{D^*})$, where
$f_{D^*}^\perp/f_{D^*}=0.9\pm0.1$ \cite{D.Me}. The form factor
$F_1$ is obtained form \cite{D.Me}
\begin{eqnarray}\label{eq6}
F_1(q^2)=\frac{f(0)}{\Big(1-q^2/m_P^2\Big)\Big[1-\sigma_1q^2/m_P^2+\sigma_2q^4/m_P^4\Big]},
\end{eqnarray}
here the $m_P$ is the $m_{B_c}$ for $B\rightarrow D$ transition.
The values of $f(0)$, $\sigma_1$ and $\sigma_2$ are as follows
\begin{eqnarray}\label{eq7}
F_1^{B\rightarrow D}:\quad f(0)=0.67,\quad \sigma_1=0.57,\quad
\sigma_2=0.
\end{eqnarray}
For the Wilson parameter $a_j$ (j = 1, \ldots, 10), we have
\begin{eqnarray}\label{eq8}
a_{2j-1}=C_{2j-1}+\frac{1}{3}C_{2j},\quad
a_{2j}=C_{2j}+\frac{1}{3}C_{2j-1},\quad j=1,2,3,4,5.
\end{eqnarray}
We used the next-to-leading logarithm in the naive dimensional
regularization (NDR) scheme for the Wilson coefficients $C_j(\mu)$
at the scale $\mu$ that are shown in Tab. \ref{tab3}.
\begin{table}[t]
\centering\caption{\label{tab3}  Wilson coefficients $C_j$ in the
NDR scheme ($\alpha=1/129$) \cite{M.Be1}.}
\begin{tabular}{|c|c|c|c|}
  \hline
   NLO    & $\mu=m_b/2$ & $\mu=m_b$ & $\mu=2m_b$\\ \hline
 $C_1$ & 1.137 & 1.081 & 1.045 \\ \hline
 $C_2$ & -0.295 & -0.190 & -0.113 \\ \hline
 $C_3$ & 0.021 & 0.014 & 0.009 \\ \hline
 $C_4$ & -0.051 & -0.036 & -0.025 \\ \hline
 $C_5$ & 0.010 & 0.009 & 0.007 \\ \hline
 $C_6$ & -0.065 & -0.042 & -0.027 \\ \hline
 $C_7/\alpha$ & -0.024 & -0.011 & 0.011 \\ \hline
 $C_8/\alpha$ & 0.096 & 0.060 & 0.039 \\ \hline
 $C_9/\alpha$ & -1.325 & -1.254 & -1.195 \\ \hline
 $C_{10}/\alpha$ & 0.331 & 0.223 & 0.144 \\ \hline
\end{tabular}
\end{table}
In this paper, we take the decay constants, quark, and meson
masses (in units of MeV) \cite{PDG1}{\setlength\arraycolsep{.75pt}
\begin{eqnarray}\label{eq9}
m_{B_c}&=&6274.9\pm0.8,\quad m_{D^*}=2010.26\pm0.05,\quad
m_{B^0}=5279.65\pm0.12,\quad m_{D^\pm}=1869.66\pm0.05,\nonumber\\
m_b&=&4180^{+40}_{-30},\quad m_d=4.67^{+0.48}_{-0.17},\quad
m_c=1270\pm20,\quad f_{D^*}=230\pm20.\quad
\end{eqnarray}}
Similarly, $\mathcal{A}_{1,2}(\bar{B}^0\rightarrow D^{*-}D^+)$ is
calculated. The decay rates corresponding to the
$\mathcal{A}(B^0\rightarrow D^{*+}D^-)$ and
$\mathcal{A}(\bar{B}^0\rightarrow D^{*-}D^+)$ amplitudes which are
defined as \cite{M.Ar1} {\setlength\arraycolsep{.75pt}
\begin{eqnarray}\label{eq10}
\Gamma(B^0\rightarrow D^{*+}D^-)&=&\Big||\mathcal {A}_1|
e^{i(\delta_1 + \phi_1)}+|\mathcal {A}_2|e^{i(\delta_2 + \phi_2)}\Big|^2,\nonumber\\
\Gamma(\bar{B}^0\rightarrow D^{*-}D^+)&=&\Big||\mathcal {A}_1|
e^{i(\delta_1 - \phi_1)}+|\mathcal {A}_2|e^{i(\delta_2 -
\phi_2)}\Big|^2.
\end{eqnarray}}
We calculated the branching fractions for the $B^0\rightarrow
D^{*+}D^-$ decay is written as
\begin{eqnarray}\label{eq11}
\mathcal{B}(B^0\rightarrow D^{*+}D^-)=\frac{\Gamma(B^0\rightarrow
D^{*+}D^-)}{\Gamma^{tot}_{B^0}},
\end{eqnarray}
here the $\Gamma^{tot}_{B^0}$ is $(4.33\pm0.01)\times10^{-13}$
GeV. The direct $CP$ violation can be expressed as \cite{I.Be.1}
{\setlength\arraycolsep{.75pt}
\begin{eqnarray}\label{eq12}
\mathcal{A}_{D^*D}^{CP}&=&\frac{\Gamma(B^0\rightarrow
D^{*+}D^-)-\Gamma(\bar{B}^0\rightarrow
D^{*-}D^+)}{\Gamma(B^0\rightarrow
D^{*+}D^-)+\Gamma(\bar{B}^0\rightarrow D^{*-}D^+)}\nonumber\\
&=&\frac {2|\mathcal{A}_2/\mathcal
{A}_1|\sin(\delta_1-\delta_2)\sin(\phi_1-\phi_2)}{1+|\mathcal
{A}_2/\mathcal{A}_1|^2+2|\mathcal
{A}_2/\mathcal{A}_1|\cos(\delta_1-\delta_2)\cos(\phi_1-\phi_2)}.
\end{eqnarray}}
We obtained the strong phases with values
$\delta_1=-90.01^{\circ}$ and $\delta_2=67.47^{\circ}$. Also, we
calculated for weak phases $\phi_1=0.03^{\circ}$ and $\phi_2=0$.\\
In the SM, $CP$ violation occurs when more than one of the
Cabibbo-Kobayashi-Maskawa (CKM) quark mixing matrix elements is
complex. Here, we use the CKM matrix elements at order $\lambda^5$
that is \cite{HFLAV1}
\begin{eqnarray}\label{eq13}
V=\left(%
\begin{array}{ccc}
  1-1/2\lambda^2-1/8\lambda^4 & \lambda & A\lambda^3(\rho-i\eta) \\
  -\lambda+1/2A^2\lambda^5[1-2(\rho+i\eta)] & 1-1/2\lambda^2-1/8\lambda^4(1+4A^2) & A\lambda^2 \\
  A\lambda^3[1-(1-1/2\lambda^2)(\rho+i\eta)] & -A\lambda^2+1/2A\lambda^4[1-2(\rho+i\eta)] & 1-1/2A^2\lambda^4 \\
\end{array}%
\right)
\end{eqnarray}
We adopt the Wolfenstein parameterization and choose the
parameters $A, \rho, \eta$ and $\lambda$ as \cite{PDG1}
\begin{eqnarray}\label{eq14}
\lambda=0.22650\pm0.00048,\quad A=0.790^{+0.017}_{-0.012},\quad
\bar{\rho}=0.141^{+0.016}_{-0.017},\quad \bar{\eta}=0.357\pm0.011,
\end{eqnarray}
with $\bar{\rho}=\rho(1-1/2\lambda^2)$ and
$\bar{\eta}=\eta(1-1/2\lambda^2)$. Therefore, the CKM matrix
elements are obtained as follows (in units of
$10^{-3}$){\setlength\arraycolsep{.75pt}
\begin{eqnarray}\label{eq15}
 V_{cb}&=&40.529,\quad  V_{cd}=-226.368 - 0.136i,\nonumber\\
 V_{tb}&=&999.179,\quad V_{td}=7.885 - 3.277i.
\end{eqnarray}}
Another type of $CP$ violation that occurs in the $B^0$ meson
decay, is the violation from interference between decay with and
without mixing (without any of the other types of $CP$ violation). We have \cite{HFLAV2}\\
\begin{eqnarray}\label{eq16}
\lambda_{D^{*}D}=\frac{q}{p}\frac{\mathcal{\bar{A}}}{\mathcal{A}}.
\end{eqnarray}
where $A(\bar{A})$ is the decay amplitude for $B^0(\bar{B}^0)$ and
$q/p$ is the ratio of the flavor contributions to the mass
eigenstates. Since the $t$ quark has more mass, only hadrons with
$c$ or $u$ quarks are allowed to transition to physical states. In
this case, we have two probability restrictions for these
transitions: first, the decay of both $B^0$ and $\bar{B}^0$ are
Cabibbo-suppressed, second, the decay for $B^0$ is
Cabibbo-allowed, and for $\bar{B}^0$ mesons doubly
Cabibbo-suppressed, or vice versa. Therefore, the decay width
difference is small compared to the mass difference, which allows
us to express $q/p$ in terms of CKM matrix elements as
\begin{eqnarray}\label{eq16}
\frac{q}{p}\approx\sqrt{\frac{M_{12}^*}{M_{12}}}=\frac{V^*_{tb}V_{td}}{V_{tb}V^*_{td}}.
\end{eqnarray}
The $M_{12}$ and $M_{12}^*$ are denote mass matrices. If
$|\lambda|\neq1$, $CP$ violation is manifest through either decay
or mixing, but if $Im\lambda\neq0$, $CP$ violation is
manifest through the interference between decays with and without mixing.\\
The different $CP$ parameters $\lambda$ arise from the fact that
the relative contribution of the penguin diagrams need not be the
same for the $0,\perp$ and $\parallel$ amplitudes \cite{BABAR3}.
The decay time-dependent $CP$ asymmetry,
$\mathcal{A}_{D^*D}^{CP}(t)$, can be defined \cite{AJ.B}
\begin{eqnarray}\label{eq17}
\mathcal{A}_{D^*D}^{CP}(t)=\frac{S_{D^{*+}D^-}\sin(\Delta
m_dt)-C_{D^{*+}D^-}\cos(\Delta m_dt)}{\cosh(\Delta\Gamma
t/2)-A_{D^{*+}D^-}^{\triangle\Gamma}\sinh(\Delta\Gamma t/2)}
\end{eqnarray}
where $\Delta m_d=0.510\hbar ps^{-1}$ and with \cite{X.Yu1}.
{\setlength\arraycolsep{.75pt}
\begin{eqnarray}\label{eq18}
S_{D^{*+}D^-}=\frac{2Im\lambda_{D^{*+}D^-}}{1+|\lambda_{D^{*+}D^-}|^2},
C_{D^{*+}D^-}=\frac{1-|\lambda_{D^{*+}D^-}|^2}{1+|\lambda_{D^{*+}D^-}|^2},
A_{D^{*+}D^-}^{\triangle\Gamma}=-\frac{2Re\lambda_{D^{*+}D^-}}{1+|\lambda_{D^{*+}D^-}|^2},
\end{eqnarray}
For them also applies \cite{X.G.1}
\begin{eqnarray}\label{eq19}
(S_{D^{*+}D^-})^2+(C_{D^{*+}D^-})^2+(A_{D^{*+}D^-}^{\triangle\Gamma})^2=1,
\end{eqnarray}
and this constraint may or may not imposed to fits. To calculate
the mixing-induced and direct $CP$ violation, we use
$S_{D^{*+}D^-}$ and $C_{D^{*+}D^-}$ parameters, respectively.
parameter $|A_{D^{*+}D^-}^{\triangle\Gamma}|$ introduces another
observable for neutral meson systems. In the $B^0$ decay, the
expression for the time-dependent amplitude
$\mathcal{A}_{D^*D}^{CP}(t)$ is simplified because of the low
oscillation frequency. Therefore, the Eq. (\ref{eq17}) becomes
\cite{Belle3}
\begin{eqnarray}\label{eq20}
\mathcal{A}_{D^*D}^{CP}(t)=S_{D^{*+}D^-}\sin(\Delta
m_dt)-C_{D^{*+}D^-}\cos(\Delta m_dt)
\end{eqnarray}
By changing the final state ($D^{*+}D^-$ to $D^{*-}D^+$), the
values of $S_{D^{*-}D^+}$, $C_{D^{*-}D^+}$ and
$A_{D^{*-}D^+}^{\triangle\Gamma}$ are obtained. From the
combination of final states $D^{*+}D^-$ and $D^{*-}D^+$, the
following $CP$ parameters for the $B^0\rightarrow D^{\pm*}D^{\mp}$
decay can be defined \cite{M.C1}
\begin{eqnarray}\label{eq21}
S_{D^*D}=\frac{1}{2}(S_{D^{*+}D^-} + S_{D^{*-}D^+}),\quad
\triangle S_{D^*D}=\frac{1}{2}(S_{D^{*+}D^-} - S_{D^{*-}D^+}), \nonumber\\
C_{D^*D}=\frac{1}{2}(C_{D^{*+}D^-} + C_{D^{*-}D^+}), \quad
\triangle C_{D^*D}=\frac{1}{2}(C_{D^{*+}D^-} - C_{D^{*-}D^+}).
\end{eqnarray}
The $S_{D^*D}$ is mixing induced $CP$ violation However $\triangle
S_{D^*D}$ is insensitive to $CP$ violation because is related to
the strong phase. In the case of $CP$ invariance,
$S_{D^{*+}D^-}=-S_{D^{*-}D^+}$ is fulfilled. The $C_{D^*D}$ is
direct $CP$ violation and $\triangle C_{D^*D}$ define the
asymmetry between the rates $\Gamma(B^0\rightarrow
D^{*+}D^-)+\Gamma(\bar{B}^0\rightarrow D^{*-}D^+)$ and
$\Gamma(B^0\rightarrow D^{*-}D^+)+\Gamma(\bar{B}^0\rightarrow
D^{*+}D^-)$ \cite{C.H.1}. The $\triangle C_{D^*D}=\pm1$ denotes a
flavour-specific decay, where no $CP$ violation in the
interference between decay and decay after mixing is feasible,
while decays with $\triangle C_{D^*D}=0$ have the highest
sensitivity to mixing induced $CP$ violation.

\section{Numerical results and conclusion}
The $CP$ parameters resulting from the fit to the decay time,
direct $CP$ violation and branching ratio for the $B^0\rightarrow
D^{*+}D^-$ decay are shown in Tab. \ref{tab4}.
\begin{table}[t]
\centering\caption{\label{tab4} The $CP$ violation parameters and
branching ratio for $B^0\rightarrow D^{*+}D^-$ decay at three
different choices of $\mu$ scale.}
\begin{tabular}{|c|c|c|c|c|}
  \hline
parameters & $\mu=m_b/2$ & $\mu=m_b$ & $\mu=2m_b$ &  Exp. \\
\hline
$S_{D^{*+}D^-}$ & $-0.630\pm0.021$ & $-0.655\pm0.021$ & $-0.673\pm0.022$ & $-0.80\pm0.09$ \cite{PDG1} \\
\hline
$C_{D^{*+}D^-}$ & $-0.025\pm0.015$ & $-0.031\pm0.016$ & $-0.029\pm0.015$ & $-0.03\pm0.09$ \cite{PDG1} \\
\hline
$S_{D^{*-}D^+}$ & $-0.734\pm0.031$ & $-0.763\pm0.031$ & $-0.748\pm0.031$ & $-0.83\pm0.09$ \cite{PDG1} \\
\hline
$C_{D^{*-}D^+}$ & $-0.036\pm0.006$ & $-0.071\pm0.008$ & $-0.080\pm0.009$ & $-0.02\pm0.08$ \cite{PDG1} \\
\hline
 $S_{D^*D}$ & $-0.707\pm0.022$ & $-0.709\pm0.024$ & $-0.710\mp0.024$ & $-0.861\pm0.077\pm0.019$ \cite{LHCb1} \\
 \hline
 $C_{D^*D}$ & $-0.030\pm0.003$ & $-0.051\pm0.004$ & $-0.054\pm0.004$ & $-0.059\pm0.092\pm0.020$ \cite{LHCb1}\\
 \hline
 $\triangle S_{D^*D}$ & $0.076\pm0.004$ & $0.054\pm0.003$ & $0.037\pm0.002$ & $+0.019\pm0.075\pm0.012$ \cite{LHCb1}\\
 \hline
 $\triangle C_{D^*D}$ & $0.005\pm0.000$ & $0.020\pm0.001$ & $0.026\pm0.002$ & $-0.031\pm0.092\pm0.016$ \cite{LHCb1}\\
 \hline
 $\mathcal{A}_{D^*D}^{CP}$ & $0.011\pm0.001$ & $0.008\pm0.001$ & $0.005\pm0.001$ & $0.008\pm0.014\pm0.006$ \cite{LHCb1}\\
 \hline
 $\mathcal{B}(\times10^{-4})$ & $4.61\pm1.11$ & $4.93\pm1.19$ & $5.20\pm1.25$ & $6.03\pm0.50$ \cite{HFLAV1}\\
\hline
\end{tabular}
\end{table}
The main our goal of the analysis of $B^0\rightarrow D^{*+}D^-$
decay was to calculate the $CP$ parameters ($S_{D^*D}$,
$C_{D^*D}$, $\triangle S_{D^*D}$, $\triangle C_{D^*D}$, and
$\mathcal{A}_{D^*D}^{CP}$). Studying decays that involve $CP$
violation is a good way to verify the theoretical principles in
the quark-flavour of
the SM.\\
The $B^0\rightarrow D^{*+}D^-$ decay, involves $b\rightarrow
c\bar{c}d$ transitions, which are CKM suppressed. The
contributions of higher-order are not Cabibbo-suppressed so the
analysis of the $B^0\rightarrow D^{*+}D^-$ decay helps to
constrain these contributions in order to distinguish them from
the effects of new physics.\\
Here we have obtained the direct $CP$ violation and parameters
$CP$ violation from interference between decay with and without
mixing. The uncertainty of the calculated parameters is due to the
mass of quarks and mesons, the decay constant and CKM matrix
elements. The most important value in the theoretical uncertainty
is related to the decay constant.\\
We have calculated the $CP$ parameters as
$\mathcal{A}_{D^*D}^{CP}=0.008\pm0.001$. Also, we have found
$S_{D^*D}=-0.709\pm0.024$, $\triangle S_{D^*D}=0.054\pm0.003$,
$C_{D^*D}=-0.051\pm0.004$ and $\triangle C_{D^*D}=0.020\pm0.001$.
From the sum of the amplitudes, we have calculated the total
amplitude and obtained comparable result with experimental value
for the branching ratio as $\mathcal{B}(B^0\rightarrow
D^{*+}D^-)=(5.20\pm1.25)\times10^{-4}$ at $\mu=2m_b$ scale.


\begin{thebibliography}{20}
\bibitem{LHCb3}
R. Aaij et al., LHCb collaboration, \textit{Measurement of CP
violation in $B^0\rightarrow D^+D^-$ decays}, Phys. Rev. Lett
\textbf{117} (2016) 261801.
\bibitem{Belle2}
T. Aushev et al., Belle Collaboration, \textit{Search for CP
violation in the decay $B^0\rightarrow D^{*\pm}D^{\mp}$}, Phys.
Rev. Lett \textbf{93} (2004) 201802.
\bibitem{AJ1}
A.J. Bevan et al., BABAR and Belle Collaborations \textit{The
Physics of the B Factories}, Eur. Phys. J. C \textbf{74} (2014)
3026.
\bibitem{PH1}
P. Huet and E. Sather, \textit{Electroweak Baryogenesis and
Standard Model CP Violation}, Phys. Rev. D \textbf{51} (1995) 379.
\bibitem{ZH1}
ZH. Xiao, W. Li, L. Guo and G. Lu, \textit{Charmless decays
$B\rightarrow PP, PV$, and effects of new strong and electroweak
penguins in Topcolor-assisted Technicolor mode}, Eur. Phys. J. C
\textbf{18} (2001) 681.
\bibitem{HFLAV2}
Y. Amhis et al., Heavy Flavor Averaging Group, \textit{Averages of
b-hadron, c-hadron, and $\tau$-lepton properties as of summer
2016}, Eur. Phys. J. C \textbf{77} (2017) 895.
\bibitem{Belle1}
 M. Rohrken et al., Belle Collaboration, \textit{Measurements of Branching Fractions and Time-dependent CP Violating Asymmetries in
 $B^0\rightarrow
 D^{(*)\pm}D^{\mp}$ Decays}, Phys. Rev. D \textbf{73} (2006) 112004.
\bibitem{BABAR1}
B. Aubert et al., BABAR Collaboration, \textit{Measurements of
time-dependent CP asymmetries in $B^0\rightarrow D^{(*)+}D^{(*)-}$
decays}, Phys. Rev. D \textbf{79} (2009) 032002.
\bibitem{LHCb1}
R. Aaij et al., LHCb collaboration, \textit{Measurement of CP
violation in $B^0\rightarrow D^{*\pm}D^{\mp}$ decays}, Journal of
High Energy Physics \textbf{03} (2020) 147.
\bibitem{HFLAV1}
Y. Amhis et al., Heavy Flavor Averaging Group, \textit{Averages of
b-hadron, c-hadron, and $\tau$-lepton properties as of 2018}, Eur.
Phys. J. C \textbf{81} (2021) 226.
\bibitem{BABAR2}
B. Aubert et al., BABAR Collaboration, \textit{Measurement of the
Branching Fraction and CP-violating Asymmetries in Neutral B
Decays to $D^{*\pm}D^{\mp}$}, Phys. Rev. Lett \textbf{90} (2003)
221801.
\bibitem{PDG1}
P.A. Zyla et al., Particle Data Group, \textit{Review of particle
physics}, Prog. Theor. Exp. Phys. \textbf{2020} (2020) 083C01.
\bibitem{A.I1}
A.I. Sanda and Z.ZH. Xing, \textit{Towards Determining $\phi_1$
with $B\rightarrow D^{(*)}\bar{D}^{(*)}$}, Phys. Rev. D
\textbf{56} (1997) 341.
\bibitem{G.B1}
G. Buchalla, \textit{CP Violation in K and B Decays}, AIP Conf.
Proc \textbf{412} (1997) 49.
\bibitem{M.S.S}
M.S. Sozzi and I. Mannelli, \textit{Measurements of direct $CP$
violation}, Riv. Nuovo Cim. \textbf{26} (2003) 110.
\bibitem{D.Me}
D. Melikhov and B. Stech, \textit{Weak form factors for heavy
meson decays: An update}, Phys. Rev. D. \textbf{62} (2000) 014006.
\bibitem{M.Be1}
M. Beneke, G. Buchalla, M. Neubert and C.T. Sachrajda, \textit{QCD
factorization in $B\rightarrow \pi K, \pi\pi$ decays and
extraction of wolfenstein parameters}, Nucl. Phys. B \textbf{606}
(2001) 245.
\bibitem{M.Ar1}
M. Artuso, E. Barberio and SH. Stone, \textit{B Meson Decays},
PMC. Physics. A \textbf{3} (2009) 3.
\bibitem{I.Be.1}
I. Bediaga and C. Gobel, \textit{Direct CP violation in beauty and
charm hadron decays}, Porg. Part. Nucl. Phys \textbf{114} (2020)
103808.
\bibitem{BABAR3}
B. Aubert et al., BABAR Collaboration, \textit{A Study of
Time-Dependent CP-Violating Asymmetries and Flavor Oscillations in
Neutral B Decays at the $\Upsilon(4S)$}, Phys. Rev. D \textbf{66}
(2002) 032003.
\bibitem{AJ.B}
A.J. Bevan, G. Inguglia and B. Meadows, \textit{Time-dependent CP
asymmetries in D and B decays}, Phys. Rev. D \textbf{84} (2011)
114009.
\bibitem{X.Yu1}
X. Yu, Z.T. Zoua and C.D. Lu, \textit{Time-dependent CP-violations
of $B(B_s)$ decays in the perturbative QCD approach}, Phys. Rev. D
\textbf{88} (2013) 054018.
\bibitem{X.G.1}
X.G. He, S.F. Li, B. Ren and X.B. Yuan, \textit{Tests for CPT sum
rule and U-spin violation in Time-dependent CP violation of
$B^0_s\rightarrow K^+K^-$ and $B^0\rightarrow \pi^+\pi^-$ }, Phys.
Rev. D \textbf{96} (2017) 053004.
\bibitem{Belle3}
K. Abe et al., Belle Collaboration, \textit{Improved Measurement
of Mixing-induced CP Violation in the Neutral B Meson System},
Phys. Rev. D \textbf{66} (2002) 071102.
\bibitem{M.C1}
M. Calvi, \textit{Measurements of CP violation in $B\rightarrow
DD$ decays}, PoS. Beauty \textbf{2019} (2020) 008.
\bibitem{C.H.1}Belle3
C.H. Cheng, \textit{Measurements of the CKM Angle $\beta/{\phi}_1$
at B Factories}, ECONFC \textbf{070512} (2007) 010.
\end{thebibliography}
\end{document}